\documentclass[twocolumn,english,prc,superscriptaddress,showpacs,preprintnumbers,amsmath,amssymb,floatfix]{revtex4}
\usepackage[T1]{fontenc}
\usepackage[latin9]{inputenc}
\usepackage{units}
\usepackage{esint}

\makeatletter

\providecommand{\tabularnewline}{\\}

\@ifundefined{textcolor}{}
{%
 \definecolor{BLACK}{gray}{0}
 \definecolor{WHITE}{gray}{1}
 \definecolor{RED}{rgb}{1,0,0}
 \definecolor{GREEN}{rgb}{0,1,0}
 \definecolor{BLUE}{rgb}{0,0,1}
 \definecolor{CYAN}{cmyk}{1,0,0,0}
 \definecolor{MAGENTA}{cmyk}{0,1,0,0}
 \definecolor{YELLOW}{cmyk}{0,0,1,0}
 }

\usepackage{dcolumn}\usepackage{bm}\usepackage{longtable}

\makeatother

\usepackage{babel}

\makeatother

\usepackage{babel}

\begin{document}

\preprint{RPC}

\title{Ratio of Isoscalar to Isovector Core Polarization \\
 for Magnetic Moments}

\author{L. Zamick}

\affiliation{Department of Physics and Astronomy, Rutgers University, Piscataway,
New Jersey 08854, USA }

\author{Y.Y. Sharon}

\affiliation{Department of Physics and Astronomy, Rutgers University, Piscataway,
New Jersey 08854, USA }

\author{S. J. Q. Robinson}

\affiliation{Department of Physics, Millsaps College, Jackson, Mississippi 32901}

\date{September 7, 2010}
\begin{abstract}
In calculations of isoscalar magnetic moments of odd-odd N=Z nuclei
it was found that for medium to heavy mass nuclei large-scale shell
model calculations yielded results which were very close to those
obtained with the much simpler single-j shell model. To understand
this we compare isoscalar and isovector core-polarization configuration-mixing
contributions to the magnetic moments-of-mirror pairs in first order
perturbation theory, using a spin dependent delta interaction.We fit
the strength of the delta interaction by looking at isovector and
isoscalar mirror pairs. We then use the same interaction to calculate
corrections due to first order core polarization of the magnrtic moments
of odd-odd nuclei. 
\end{abstract}
\maketitle
Previously S. Yeager et al.\cite{yeager09} compared the results of
both large-scale and single-j shell model calculations of isoscalar
magnetic g factors of odd-odd N=Z nuclei with the experimental data.
There it was noted that for large A the g factors, both experimental
and calculated, approached the value of 0.5. Detailed comparisons
with experiment were made in \cite{yeager09}. In the present work
the emphasis will be on a theory vs. theory discussion of these results,
based on the observation there that as A increases the results of
the large-scale shell-model calculations for these odd-odd N=Z nuclei,
were becoming very close to the single-j shell-model results. This
is shown in Table 1. We see that for light nuclei, e.g. $^{6}$Li,
there are significant differences between the results of the two models.
However, the differences become much smaller as A increases.

%
\begin{table}
\caption{Comparison of large-scale shell model and single-j shell model results
for Isoscalar g factors of N=Z odd-odd nuclei}

\label{tab:compare} \begin{tabular}{|c|c|c|c|c|}
\hline 
Nucleus  & J$^{\pi}$  & Large scale  & Single-j  & single-j values \tabularnewline
\hline 
$^{2}$H  & 1$^{+}$  & 0.88  & 0.88  & s$_{1/2}$\tabularnewline
\hline 
$^{6}$Li  & 1$^{+}$  & 0.87  & 0.63  & p$_{3/2}$\tabularnewline
\hline 
$^{10}$B  & 3$^{+}$  & 0.61  & 0.63  & p$_{3/2}$\tabularnewline
\hline 
$^{14}$N  & 1$^{+}$  & 0.32  & 0.37  & p$_{1/2}$\tabularnewline
\hline 
$^{18}$F  & 3$^{+}$  & 0.62  & 0.58  & d$_{5/2}$\tabularnewline
\hline 
 & 5$^{+}$  & 0.58  & 0.58  & d$_{5/2}$ \tabularnewline
\hline 
$^{22}$Na  & 3$^{+}$  & 0.59  & 0.58  & d$_{5/2}$\tabularnewline
\hline 
 & 1$^{+}$  & 0.52  & 0.58  & d$_{5/2}$\tabularnewline
\hline 
$^{26}$Al  & 5$^{+}$  & 0.57  & 0.58  & d$_{5/2}$\tabularnewline
\hline 
$^{38}$K  & 3$^{+}$  & 0.41  & 0.42  & d$_{3/2}$\tabularnewline
\hline 
$^{46}$V  & 3$^{+}$  & 0.58  & 0.55  & f$_{7/2}$\tabularnewline
\hline 
$^{58}$Cu  & 1$^{+}$  & 0.63  & 0.63  & p$_{3/2}$\tabularnewline
\hline
\end{tabular}
\end{table}


Overall, the average absolute deviation between the two columns for
nuclei heavier than $^{6}$Li is about 0.02 (i.e. about 4\%).

A related problem is that of the magnetic moments of mirror pairs.
We can form isoscalar and isovector combinations defined thus:

\begin{equation}
\mu(IS)=(\mu(oddproton)+\mu(oddneutron))/2\end{equation}

\[
\mu(IV)=(\mu(oddproton)-\mu(oddneutron))/2\]

In this work we will focus on the mirror pairs $^{57}$N, $^{57}$Cu
and the odd-odd nucleus $^{58}$Cu. Our plan of attack is to use a
spin dependent surface delta interaction -G(1+xP$_{\sigma}$) delta($\vec{r}_{1}-\vec{r}_{2}$)
to calculate deviations from the Schmidt limits for the above 3 nucleii.
Only one parameter G enters into the calculaiton for $\mu$(IV) and
hence we will use this moment to fit G. We will then examine the values
of the spin exchange parameter needed to explain the isoscalar magnetic
moments i.e. $\mu$(IS) and $\mu$(58Cu). Indeed, the main thrust
of this work will be to show the importance of the spin exchange in
explaining why the isoscalar moments are so close to the single-j(Schmidt)
results.

For the mirror pairs we have the follwing information for the J=3/2$^{-}$
ground states,where we give the values in units of nuclear magnetons:

$\mu$$^{57}$Ni)=-0.7975 from T. Ohtsubo et al.{[}6{]}.

$\mu^{57}$Cu)=2.582(7) from T.E. Cocolios et al. {[}7{]}.

The latter is quite different from a previous value of 2.0 in Ref
{[}8{]}.

From the above we find, using (1)

$\mu$(IS) =0.892 ,$\mu$(IV)=1.690. The Schmidt values are 0.940
and 2.853 respectively.The deviations of the magnetic moments (expt.-theor.)
are respectively -.0480 and -1.163.

The deviations of the g factors are (expt.-theor.) are respectively
-0.0320 and -0.775. An good estimate of the deviation from the Schmidt
value for $^{58}$Cu (with J=1) is the above isoscalar value -0.03184.Unfortunately
the error bars for the measurement on this nucleus are too large.

The ratio of deviations (IS/IV) is 0.04256 i.e.the isoscalar deviation
is much smaller than the isovector one.

To understand why this is so we introduce configuration mixing via
perturbation (PT) theory following Arima and his group \cite{arima54,noya58}.
We thus calculate the core-polarization configuration-mixing corrections
to these magnetic moments that are calculated in the single-j shell
model. The PT expression for these corrections simplifies if we use
particle-hole, rather than particle-particle, interaction matrix elements
(see e.g. Eq.5 in Mavromatis et al. \cite{mavro67}). This is because
for each isospin there is only one particle-hole matrix element with
total angular momentum one, while there are several such particle-particle
matrix elements. The expression is as follows.

\begin{equation}
\Delta\mu(T)=2(j/((j+1)2j+1)))^{1/2}*V(ph,T)*<j'||\mu(T)||j>/(\epsilon_{j}-\epsilon_{j'})\end{equation}

\[
whereV(ph,T)=<(j_{v}j_{v}^{-1})J=1,T|V|(j^{'}j^{-1})J=1,T>\]

Here j$_{v}$ is the angular momentum of the odd particle,j of the
core particle that is excited to the state j', T is the isospin of
the particle-hole pair.In the case here considered j$_{v}$ is $p{}_{3/2}$,
j is $f_{7/2}$ and j' is $f_{5/2}$. The magnetic moment operator
can only connect to the same state or to spin-orbit partner.

As mentioned above, we determine, the strength of the interaction,
by looking at the mirror pairs nuclei from which we can form isoscalar
and isovector combinations.Since the isoscalar corrections are very
smal we use the isovector ones to determine the spin independent part
of the interaction.

The expression given in \cite{mavro67} can be used to give, in first
order, the ratio $\frac{\Delta\mu(0)}{\Delta\mu(1)}$ of the isoscalar
to isovector core-polarization corrections to the magnetic moments
that were calculated in the single-j shell model ..

This ratio is: \begin{equation}
\frac{\Delta\mu(0)}{\Delta\mu(1)}=[\frac{V(ph,0)}{V(ph,1)}]*[\frac{<j'||\vec{\mu(0)}||j>}{<j'||\vec{\mu(1)}||j>}]\end{equation}
 In taking this ratio many factors cancel out.

In the above expressions we have

\begin{equation}
isoscalar:\indent\vec{\mu}(0)=0.5\vec{L}+0.88\vec{S}\end{equation}
 and \begin{equation}
isovector:\indent\vec{\mu}(1)=0.5\vec{L}+4.71\vec{S}\end{equation}

These results follow from the g factors of the free nucleaons and
the definitions (see eq.(0))

\begin{equation}
\vec{\mu}_{p}\equiv\vec{\mu}(0)+\vec{\mu}(1)\indent\vec{\mu}_{n}\equiv\vec{\mu}(0)-\vec{\mu}(1)\end{equation}

We now use for V in our calculations a delta interaction

V(r)= -G $\delta$($\vec{r}$)

(1+ x P$^{\sigma}$) where P$^{\sigma}$ is the spin exchange operator
(1+$\sigma$$_{1}$.$\sigma$$_{2}$)/2

The expression for the particle-hole matrix elements with this delta
interaction is {[}6{]}

V(ph,0)= G/2 $\bar{R}$(1-2x) M

V(ph,1)=G/2 $\bar{R}$M

where M= (2j$_{v}$+1) ((2j+1$)(2j$'+1))$^{1/2}${[}$(\begin{array}{ccc}
j_{v} & j_{v} & 1\\
1/2 & -1/2 & 0\end{array}$) $(\begin{array}{ccc}
j' & j & 1\\
1/2 & -1/2 & 0\end{array}$) + ($\begin{array}{ccc}
j_{v} & j_{v} & 1\\
1/2 & 1/2 & -1\end{array}$) $\begin{array}{ccc}
j' & j & 1\\
1/2 & 1/2 & -1\end{array}$){]}

$\bar{R}=\frac{1}{4\pi}\int R_{1}(r)R_{2}(r)R_{3}(r)R_{4}(r)r^{2}dr$

We then find (V(ph,0)/V(ph,1)) = (1-2x).

A popular choice for x is 1/3. With this choice (1-2x) is equal to
1/3, i.e. the first bracket in Eq. (2) makes the ratio IS/IV=1/3 for
the core polarization corrections. But there is also the second bracket
in Eqn. (2) which we now evaluate.

Since $\vec{L}=\vec{J}-\vec{S}$, \begin{equation}
\vec{\mu}=g_{L}\vec{L}+g_{S}\vec{S}=g_{L}\vec{J}+(g_{S}-g_{L})\vec{S}\end{equation}

$\vec{\mu}$ can thus connect only states with the same L value, i.e.
the spin-orbit partners J=L+1/2 and J=L-1/2.

Then \begin{equation}
<j'||\vec{\mu}||j>=<j'|g_{L}\vec{J}|j>+(g_{S}-g_{L})<j'|\vec{S}|j>\end{equation}

The first term vanishes since $j\ne j'$ so $<j'||\vec{\mu}||j>\sim(g_{S}-g_{L})$.

All in all we find $\Delta\mu(IS)/\Delta\mu(IV)$= (1-2x){*}(g$_{S}$-g$_{L}$)$_{IS}$/(g$_{S}$-g$_{L}$)$_{IV}$

For the second bracket in Eq. (2),we need to evaluate the ratio of
$(g_{S}-g_{L})$ IS / $(g_{S}-g_{L})$ IV.

The relevant values are \cite{yeager09}

Isoscalar: \ \ $g_{L}=0.5$, \ \ $g_{S}=0.88$, \ ($g_{S}-g_{L}$)
= 0.38

Isovector:\ \ $g_{L}=0.5$, \ \ $g_{S}=4.71$, \ ($g_{S}-g_{L)}$=4.21.

This leads to a ratio of $(g_{S}-g_{L})$ IS / $(g_{S}-g_{L})$ IV
= 0.38/4.21 $\approx$ 0.09. From eq(2) we find

$\Delta\mu(1)=\frac{+8.664G\bar{R}}{\Delta E}$. To get the g factor
we divide this by 1.5(that would be $\Delta\mu$ for the J=1$^{+}$
state of $^{58}$Cu).

If we take $\Delta E$=$\epsilon_{f_{7/2}}-\epsilon_{f_{5/2}}=-5.0$
MeV and if we adjust G$\bar{R}$ so that $\mu(IV)-\mu(IV)_{Schmidt}=-1.123$
we obtain GR=0.648.

Then we ask what value in the spin exchange operaror will yield the
isoscalar difference $\mu$(IS)-$\mu$(IS)$_{Schmidt}$=-0.0478. We
thus find that this value of x is 0.264.This is to be compared with
x=1/3 for which the coupling of T=0 in the particle-particle channel
is twice that for T=1 (i.e. (1+x)/(1-x)=2).

Alternately we can use renormalized values of g$_{S}$ and g$_{L}$
due to second order perturbation theory andcorrections and meson exchange
currents. A simple choice is to only change the isovector couplings--make
g$_{S}$(IV)=0.7 {*}gs(IV)$_{free}$ and change g$_{L}$(IV) from
0.5 to o.6. These changes will make the isovector contribution smaller.
We now will have

g$_{S}$(IV)-g$_{L}$(IV)=2.697 (as compared with the free value of
4.21).The new value of x is a somewhat larger 0.349.

The main point of this paper is to show how the spin exchange term
in the nucleon-nucleon interaction affects isoscalar and isovector
moments. The (1-2x) factor in the T=0 particle-hole channel supresses
the isoscalar magnetic moment deviations from the Schmidt value by
about a factor of two or three relative to the isovector ones. 

We briefly add that in the single j shell of both neutrons and protons
the isoscalar g factor is the same as the Schmidt. -not so in isovector
case .This is relevent to the case of $^{46}$V J=3$^{+}$and the
J=$\nicefrac{7}{2}^{-}$ mirror pairs $^{45}$Ti and $^{45}$V. The
isoscalar g factor is

\[
g(IS)=\frac{(g_{j_{\pi}}+g_{\nu})}{2}\]

It does not depend on the details of the function.

The isovector term is

$[g(^{45}V)-g(^{45}Ti)]/2$

is given by

$\frac{(g_{j_{\pi}}-g_{j_{\nu}})}{2}\sum\frac{|D(J_{p}J_{n})|^{2}[J_{p}(J_{p}+1)-J_{n}(J_{n}+1)]}{J(J+1)}$

where $|D|^{2}$ is the probability that in a state of total angular
momentum J the protons couple to J$_{p}$ and the neutrons to J$_{n}$.Only
if the even number of nucleons would couple to zero would we get the
Schmidt value in the isovector case.Unfortunately the magnetic moment
of $^{45}$V has not yet been measured. 

We thank Justin Faris and Diego Torres for their assistance.SJQR acknowledges
a Millsaps College Faculty Development Award. YYS thanks Stockton
college for a Sabbatical Grant.


\begin{thebibliography}{11}
\bibitem{yeager09} S. Yeager, S. J.$ $ Q. Robinson, L. Zamick, and
Y. Y. Sharon, EPL 88, 52001 (2009)..

\bibitem{arima54} A. Arima and H. Horie, Prog. Theor. Phys. 11, 509
(1954); ibid 12, 623 (1954).

\bibitem{noya58} H. Noya, A. Arima and H. Horie, Suppl. Prog. Theor.
Phys. 8, 331 (1958).

\bibitem{mavro67} H. A. Mavromatis and L. Zamick, Nucl. Phys. A104,
17 (1967) Eq 5.

\bibitem{abbas80} A. Abbas and L. Zamick, PRC 22, 1755 (1980).

\bibitem{key-1}T. Ohtsubo et al. Phys.Rev.C54,544(1996)

\bibitem{key-2}T.E. Cocolios et al.,Phys. Rev. Lett.103,102501 (2009)

\bibitem{key-3}K. Minamisono et al., Phys. Rev. Lett. 96,102501 (2006)

\bibitem{key-1}N.J.Stone et.al,Phys. Rev C77,067302 (2008)

\bibitem{last1} J. Kramer et. al. Physics Letters B678, 465 (2008).

\bibitem{last2} K. Sugimoto, Phys. Rev. 182, 1051 (1969). 
\end{thebibliography}
\end{document}